\documentclass[useAMS,usenatbib]{mn2e}
\usepackage{graphicx}

\title[Spectral indices of Galactic radio loops]
{Spectral indices of Galactic radio loops between 1420, 820 and 408
MHz}

\author[V. Borka]{V. Borka\thanks{E-mail: vborka@vin.bg.ac.yu} \\
Laboratory of Physics (010), Vin\v ca Institute of Nuclear Sciences, P.
O. Box 522, 11001 Belgrade, Serbia }

\begin{document}

\date{Accepted 2006 December 27. Received 2006 December 17; in original form 2005 December 7}

\pagerange{1--10}
\pubyear{2007}

\maketitle

\label{firstpage}

\begin{abstract}
In this paper the average brightness temperatures and surface
brightnesses at 1420, 820 and 408 MHz of the six main Galactic
radio-continuum loops are derived, as are their radio spectral
indices. The temperatures and surface brightnesses of the radio
loops are computed using data taken from radio continuum surveys at
1420, 820 and 408 MHz. We have demonstrated the reality of Loops V
and VI and present diagrams of their spectra for the first time. We
derived the radio spectral indices of Galactic radio loops from
radio surveys at three frequencies (1420, 820 and 408 MHz) and
confirm them to be non-thermal sources. Diameters and distances of
Loops I-VI were also calculated. The results obtained are in good
agreement with current theories of supernova remnant (SNR) evolution
and suggest that radio loops may have a SNR origin.
\end{abstract}


\section{Introduction}

The fact that some radio spurs can be joined into small circles is
very well known. The sets of spurs which form small circles are
called loops. During the early seventies, four major loops were
recognized. Their discoveries and studies took place in this order:
Loop I (\citet{larg62}; \citet{hasl64}; \citet{larg66};
\citet{salt70}), Loop II (\citet{larg62}; \citet{quig65};
\citet{salt70}), Loop III (\citet{quig65}; \citet{salt70}) and Loop
IV (\citet{larg66}; \citet{salt70}; \citet{reic81}). Salter in 1970.
gave the most precise determination of these circles' parameters,
which were later published in \citet{berk71a}. The data Salter used
were the best available at the time at 408, 404, 240 and 178 MHz. A
detailed review of the subject was published by \citet{salt83}.
\citet{jmt72, jmt82} and \citet{jmt97} made an observation that some
other spurs could be connected into loops. They proposed Loop V to
be formed by negative latitude spurs in Taurus, Pisces and Pegasus
and Loop VI to correspond to the weak positive latitudes spurs in
Leo and Cancer. Using the survey at 38 MHz by \citet{jmt73} they
computed parameters of the proposed loops.

In order to study the structure of Galactic radio loop emission, it
is necessary to determine their spectral indices. The differential
spectral indices of the North Polar Spur (NPS), the spurs in
Aquarius, Pegasus and Taurus were derived using data at 1420, 408
and 38 MHz \citep{jmt95a, jmt95b}. The technique used was that of
temperature-temperature plots ($T-T$ plots). \citet{berk71} gives
indices for the spurs seen on her survey. \citet{jmt82} derived
differential temperature spectral indices between 404 and 38 MHz.

Spectral index variations of the Galactic radio continuum emission
were discussed in several papers. \citet{laws87} has studied these
variations across the northern sky between 1420 and 38 MHz and
suggested that the obtained values (about 2.5) are mainly due to the
effects of Loops I and III. Also, \citet{reic88a} used radio
continuum surveys at 1420 and 408 MHz to calculate a map of spectral
indices across the northern sky and found significant variations
between 2.3 and 3.0. The steepest spectra corresponded to the North
Polar Spur and Loop III. The same authors \citep{reic88b} gave the
estimation $\beta > 2.8$ for Loops I and III which are believed to
be nearby, very old supernova remnants. These estimates were mostly
derived using the method of $T-T$ plots over areas of the sky where
no strong local features are superposed on the general Galactic
background. \citet{webs74} studied the spectrum of the galactic
non-thermal background radiation using observations at 408, 610 and
1407 MHz and found that the mean differential temperature spectral
index was close to 2.8.

With the aim of studying spectral indices of the major radio loops
I-IV, as well as of the radio loops V and VI, we used radio
continuum surveys at three frequencies: 1420 MHz \citep{reic86}, 820
MHz \citep{berk72}, 408 MHz \citep{hasl82} and calculated the
corresponding mean temperatures and brightnesses. Knowing three
values of brightnesses, the spectral indices could be calculated as
coefficients of linear fits to logaritmic temperature versus
frequency plots. Until now, there were no generated spectra using
mean temperatures for (at least) two different frequencies, as is
necessary for forming the simplest linear kind of spectrum. All
earlier determinations of the radio loop spectral indices were based
on $T-T$ methods. The reality of Loops I-VI is investigated by
comparing the values of their spectral indices with earlier
mentioned studies of spectral indices distribution across the
northern sky. In addition, diameters and distances of Loops I-VI were
calculated using the $\Sigma - D$ relation for SNRs given by
\citet{case98}.

\section{Data}

The surveys at 1420 MHz \citep{reic86} and 820 MHz \citep{berk72}
and at 408 MHz \citep{hasl82} provided the database for the
recomputing of small circles for the major loops. The data were
obtained from the 1420-MHz  Stockert survey \citep{reic86}, the
820-MHz Dwingeloo survey \citep{berk72} and the 408-MHz all-sky
survey \citep{hasl82}. The angular resolutions are 35$'$ and 1$^\circ$.2
and 0$^\circ$.85, respectively. The effective sensitivities are about
50 mK T$_\mathrm{b}$ (T$_\mathrm{b}$ is for an average brightness
temperature) at 1420 MHz, 0.20 K at 820 MHz, and about 1.0 K at 408
MHz. These data are available$_{}$ on MPIfR's Survey Sampler (''Max
-- Planck -- Institut f\"{u}r Radioastronomie'', Bonn). This is an
on-line service (http://www.mpifr-bonn.mpg.de/survey.html), which
allows users to pick a region of the sky and obtain images at a
number of wavelengths.

\section{Analysis}

The 1420, 820 and 408 MHz continuum surveys specified in Section 2
were used to derive brightness temperature vs frequency spectra, and
the spectral indices of Loops I-VI were derived from these.

The average brightness temperatures of the Loops at the three
frequencies were estimated from the survey data. The areas of the
loops were divided into different sections (corresponding to spurs)
and estimates for these sections were combined. The longitude and
latitude ranges which include spurs of the Loops I-VI are given in
Table \ref{tab01}. The areas inside the spurs were chosen using the
three surveys specified above. Background radiation was subtracted
in this way. First, the temperature of the loop plus background was
determined. Next, the background alone near the loop was estimated.
Finally, we calculated the difference of these values to yield the
average temperature of the loop.

Average 1420, 820 and 408 MHz brightness temperatures and surface
brightnesses derived assuming the Rayleigh-Jeans approximation to
hold, are presented in Table \ref{tab02}.

Assuming the spectra to have a power-law form:

\begin{equation}
T_\mathrm{b} = K \nu ^{-\beta}
\end{equation}

\noindent
we get:

\begin{equation}
\log T_\mathrm{b} = \log K -\beta  \log \nu
\end{equation}
\noindent where $\beta $ is spectral index and K is a constant.
Knowing three values of brightnesses (derived in this paper) we
were able to derive spectral indices from fitting equation (2) to
the data. The results are given in Table \ref{tab03}.

Temperatures at 1 GHz were calculated by use of the spectral
indices:

\begin{equation}
T_\mathrm{1GHz} / T_\mathrm{\nu GHz} = (1. / \nu) ^{-\beta}
\end{equation}

\noindent We have converted these values that have been
extrapolated to 1 GHz into surface brightness and present them in
Table \ref{tab04}.

\section{Method of calculation}

The areas used to represent the individual loops were obtained from
the radio continuum map of \citet{reic86} in such a way that they
contain the brightness temperature contours enclosing the spur
features that define the loop. The longitude and latitude ranges of
each are given in Table \ref{tab01}. Contour lines which correspond
to the minimum and maximum brightness temperatures for each spur,
are taken to define their borders. T$_\mathrm{min}$ is the lower
temperature limit between the background and the spur and
T$_\mathrm{max}$ is the upper temperature limit between the spur and
unrelated confusing sources (superposed on the spur and hence
requiring elimination from the calculation). In this manner
background radiation was considered as radiation that would exist if
there were no spurs. We used averages over the data within these two
curves: the contour for T$_\mathrm{min}$ and the contour for
T$_\mathrm{max}$.

For evaluating brightness temperatures of the background, we used
all measured values below T$_\mathrm{min}$, inside the corresponding
intervals of Galactic longitude (l) and Galactic latitude (b), and
lying on the outer side of a spur. The value of T$_\mathrm{b}$ is
approximately constant near a spur. For evaluating the brightness
temperatures of a loop including the background, we used all
measured values between T$_\mathrm{min}$ and T$_\mathrm{max}$ inside
the corresponding regions of l and b. If the value of
T$_\mathrm{min}$ (or T$_\mathrm{max}$) is changed by a small amount,
say approximately $2 \times \Delta T$ (i.e. 0.1 K for 1420 MHz, 0.4
K for 820 MHz and 2 K for 408 MHz), the brightness contours become
significantly different. If T$_\mathrm{min}$ is too small, the area
of the spur becomes confused with the background and it becomes
obvious that the border has been incorrectly chosen. If
T$_\mathrm{max}$ is too large, the area of the spur includes
significant contributions from unrelated confusing sources. Mean
brightness temperatures for spurs are found by subtracting the mean
values of background brightness temperature from the mean values of
the brightness temperature over the areas of the spurs.

We calculated the mean background levels over regions not too far
outside the spurs because the loops are moving through regions of
the interstellar medium where earlier supernova remnants are likely
to have passed, i.e. through a low density medium produced by
earlier SNRs (or other energetic phenomena) \citep{mcke77, salt83}.

We used averages over the defined areas and subtracted from these
the average values for the background regions obtained as
described above.

\begin{table*}
\centering
\caption{The Galactic longitude and latitude for spurs
belonging to Loops I-VI and the lower and upper temperature limits
for these loops.}
\begin{tabular}{|l|l|l|c|c|c|c|c|c|}
\hline
\raisebox{-1.50ex}[0cm][0cm]{label of}&
\multicolumn{2}{|l|}{\raisebox{-1.50ex}[0cm][0cm]{~~~~~~~~l,b intervals}}&
\multicolumn{2}{|c|}{\textbf{1420 MHz}} &
\multicolumn{2}{|c|}{\textbf{820 MHz}} &
\multicolumn{2}{|c|}{\textbf{408 MHz}}  \\
\raisebox{-1.50ex}[0cm][0cm]{the loop}&
\multicolumn{2}{|l|}{\raisebox{-1.50ex}[0cm][0cm]{~~~~~~~~for spurs ($^\circ$)}}&
\multicolumn{2}{|c|}{temperature limits (K)} &
\multicolumn{2}{|c|}{temperature limits (K)} &
\multicolumn{2}{|c|}{temperature limits (K)}  \\
\cline{4-9}
 &
\multicolumn{2}{|l|}{} & T$_\mathrm{min}$& T$_\mathrm{max}$&
T$_\mathrm{min}$& T$_\mathrm{max}$& T$_\mathrm{min}$& T$_\mathrm{max}$ \\
\hline
\hline
\raisebox{-1.50ex}[0cm][0cm]{\textbf{Loop I}}&
l = [40, 0]&
b = [18, 78]&
\raisebox{-1.50ex}[0cm][0cm]{3.6}&
\raisebox{-1.50ex}[0cm][0cm]{4.1}&
\raisebox{-1.50ex}[0cm][0cm]{7.8}&
\raisebox{-1.50ex}[0cm][0cm]{9.5}&
\raisebox{-1.50ex}[0cm][0cm]{30.0}&
\raisebox{-1.50ex}[0cm][0cm]{40.0} \\
 & l = [360, 327]& b = [67, 78]& & & & & &  \\
\hline
\raisebox{-1.50ex}[0cm][0cm]{\textbf{Loop II}}&
l = [57, 30]&
b = [-50, -10]&
\raisebox{-1.50ex}[0cm][0cm]{3.6}&
\raisebox{-1.50ex}[0cm][0cm]{4.2}&
\raisebox{-1.50ex}[0cm][0cm]{7.2}&
\raisebox{-1.50ex}[0cm][0cm]{9.9}&
\raisebox{-1.50ex}[0cm][0cm]{24.0}&
\raisebox{-1.50ex}[0cm][0cm]{36.0} \\
 & l = [195, 130]& b = [-70, -2]& & & & & &  \\
\hline
\raisebox{-3.00ex}[0cm][0cm]{\textbf{Loop III}}&
l = [180, 135]&
b = [2, 50]&
\raisebox{-3.00ex}[0cm][0cm]{3.4}&
\raisebox{-3.00ex}[0cm][0cm]{4.2}&
\raisebox{-3.00ex}[0cm][0cm]{6.4}&
\raisebox{-3.00ex}[0cm][0cm]{10.0}&
\raisebox{-3.00ex}[0cm][0cm]{21.0}&
\raisebox{-3.00ex}[0cm][0cm]{38.0} \\
 & l = [135, 110]& b = [40, 55]& & & & & &  \\
 & l = [110, 70]& b = [6, 50]& & & & & &  \\
\hline
\textbf{Loop IV}&
l = [325, 285]&
b = [55, 72]&
3.58&
4.0&
7.45&
7.95&
27.5&
30.0 \\
\hline
\raisebox{-3.00ex}[0cm][0cm]{\textbf{Loop V}}&
l = [189, 178]&
b = [-25, -13]&
\raisebox{-3.00ex}[0cm][0cm]{3.6}&
\raisebox{-3.00ex}[0cm][0cm]{3.8}&
\raisebox{-3.00ex}[0cm][0cm]{7.1}&
\raisebox{-3.00ex}[0cm][0cm]{7.8}&
\raisebox{-3.00ex}[0cm][0cm]{24.0}&
\raisebox{-3.00ex}[0cm][0cm]{33.5} \\
 & l = [147, 133]& b = [-50, -39]& & & & & &  \\
 & l = [90, 80]& b = [-39, -24]& & & & & &  \\
\hline
\raisebox{-1.50ex}[0cm][0cm]{\textbf{Loop VI}}&
l = [215, 205]&
b = [29, 40]&
\raisebox{-1.50ex}[0cm][0cm]{3.38}&
\raisebox{-1.50ex}[0cm][0cm]{3.58}&
\raisebox{-1.50ex}[0cm][0cm]{5.8}&
\raisebox{-1.50ex}[0cm][0cm]{6.9}&
\raisebox{-1.50ex}[0cm][0cm]{15.0}&
\raisebox{-1.50ex}[0cm][0cm]{23.5} \\
 & l = [207, 196]& b = [6, 32]& & & & & &  \\
\hline
\end{tabular}
\label{tab01}
\end{table*}

The areas over which an average brightness temperature is determined
at each of the three frequencies are taken to be as similar as
possible within the limits of measurement accuracy. Moreover, in
some cases (e.g. at 1420 MHz for Loops IV and VI), we used values of
T$_\mathrm{min}$ and T$_\mathrm{max}$ which are below this limit
(see Table \ref{tab01}) in order to make the corresponding areas the
most comparable. However, some differences between these areas still
remain and we think that the major causes of differing borders
between the three frequencies are small random and systematic errors
in the calibrated data.

\section{Results}

The results for the six loops are presented in Tables \ref{tab02}
and \ref{tab03}. Table \ref{tab02} shows calculated average
brightness temperatures and surface brightnesses at 1420, 820 and
408 MHz. These results are in good agreement with results obtained
by \citet{berk73}, only that the present results are obtained with
more sample points. As one can see from Table \ref{tab01}, the first
two regions of Loop V are located inside the area of Loop II. The
results corresponding to these two loops, presented in Table
\ref{tab02}, are calculated out of their overlapping regions.
However, we also performed calculations for complete areas of Loops
II and V (including overlapping regions) and obtained practically
the same results for Loop II. In the case of the whole Loop V, the
results for $T_\mathrm{b}$ are slightly different from those given
in Table \ref{tab02}: 0.14 K at 1420 MHz, 0.51 K at 820 MHz and 5.1
K at 408 MHz. This difference is probably caused by superposition of
radiation from both loops. Spectra are presented in Figs 1 -- 6,
with Table \ref{tab03} giving the spectral indices computed for
Loops I-VI.

\begin{table*}
\centering
\caption{Temperatures (K) and brightnesses
(10$^{-22}$~W\,m$^{-2}$\,Hz$^{-1}$\,Sr$^{-1}$) of the radio loops
at 1420, 820 and 408 MHz, respectively.}
\begin{tabular}
{|l|c|c|c|c|c|c|}
\hline
label of &
\multicolumn{2}{|c|}{\textbf{1420 MHz}} &
\multicolumn{2}{|c|}{\textbf{820 MHz}} &
\multicolumn{2}{|c|}{\textbf{408 MHz}}  \\
\cline{2-7}
the loop & temperature& brightness&
temperature& brightness&
temperature& brightness \\
\hline
\hline
\textbf{Loop I}&
0.27 $\pm $ 0.05&
1.69 $\pm $ 0.30&
1.35 $\pm $ 0.20&
2.78 $\pm $ 0.40&
8.4 $\pm $ 1.0&
4.31 $\pm $ 0.50 \\
\hline
\textbf{Loop II}&
0.22 $\pm $ 0.05&
1.36 $\pm $ 0.30&
1.10 $\pm $ 0.20&
2.26 $\pm $ 0.40&
8.0 $\pm $ 1.0&
4.07 $\pm $ 0.50 \\
\hline
\textbf{Loop III}&
0.30 $\pm $ 0.05&
1.84 $\pm $ 0.30&
1.39 $\pm $ 0.20&
2.87 $\pm $ 0.40&
8.5 $\pm $ 1.0&
4.34 $\pm $ 0.50 \\
\hline
\textbf{Loop IV}&
0.08 $\pm $ 0.05&
0.48 $\pm $ 0.30&
0.53 $\pm $ 0.20&
1.09 $\pm $ 0.40&
3.0 $\pm $ 1.0&
1.52 $\pm $ 0.50 \\
\hline
\textbf{Loop V}&
0.13 $\pm $ 0.05&
0.80 $\pm $ 0.30&
0.58 $\pm $ 0.20&
1.19 $\pm $ 0.40&
5.6 $\pm $ 1.0&
2.86 $\pm $ 0.50 \\
\hline
\textbf{Loop VI}&
0.12 $\pm $ 0.05&
0.71 $\pm $ 0.30&
0.62 $\pm $ 0.20&
1.29 $\pm $ 0.40&
4.3 $\pm $ 1.0&
2.20 $\pm $ 0.50 \\
\hline
\end{tabular}
\label{tab02}
\end{table*}

The values of brightness obtained at different frequencies can be
inter-compared after using the spectral indices of Table \ref{tab03}
to reduce them to the values expected at 1 GHz. The good agreement
shown by such an inter-comparison is obvious (Table \ref{tab04}).

When comparing these results with values calculated at other
frequencies, and when all results for temperatures are estimated at
1 GHz, the good agreement of these results is obvious (Table
\ref{tab04}). The spectra presented in Figs 1 - 6 are good
approximations to simple power-law (straight) spectra between 408,
820 and 1420 MHz, in all but one case, that of Loop IV. The reason
for this might be the large relative error of the estimated
brightness temperature. A number of effects \citep{pach70} may play
some role in shaping the spectra of the loops at low frequency. From
analysis of differential spectral indices over a range of
frequencies, it has been concluded that the spectra of radio loops
are curved at low frequencies \citep{jmt95a,jmt95b}. From Figs 1 --
6 it can be seen that there is little spectral curvature between 408
and 1420 MHz, and that the data agree well with a linear fit. This
is the first spectral investigation of Loops V and VI to be
performed in this way. The spectral indices obtained for each of the
loops are $> 2.6$, confirming that their radio emissions all have a
non-thermal origin.

\begin{figure}
\centering
\includegraphics[width=0.45\textwidth]{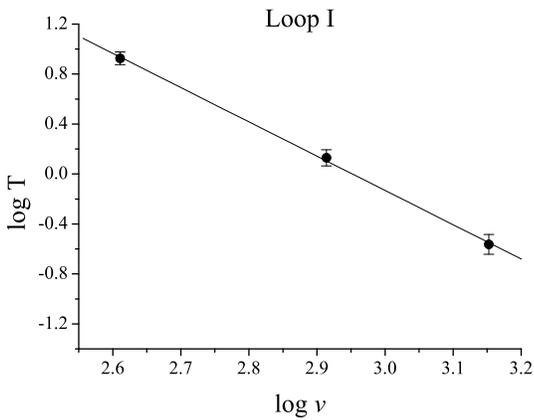}
\caption{Loop I spectrum: temperature versus  frequency, for three
measurements -- at 408, 820 and 1420 MHz. Relative errors of the
measurements $\Delta \log T = \frac{{\Delta T}}{{T \ln 10}}$ are
presented by error bars, where $\Delta T$ are the corresponding
absolute errors given in Table $\ref{tab02}$.}
\label{fig01}
\end{figure}

\begin{figure}
\centering
\includegraphics[width=0.45\textwidth]{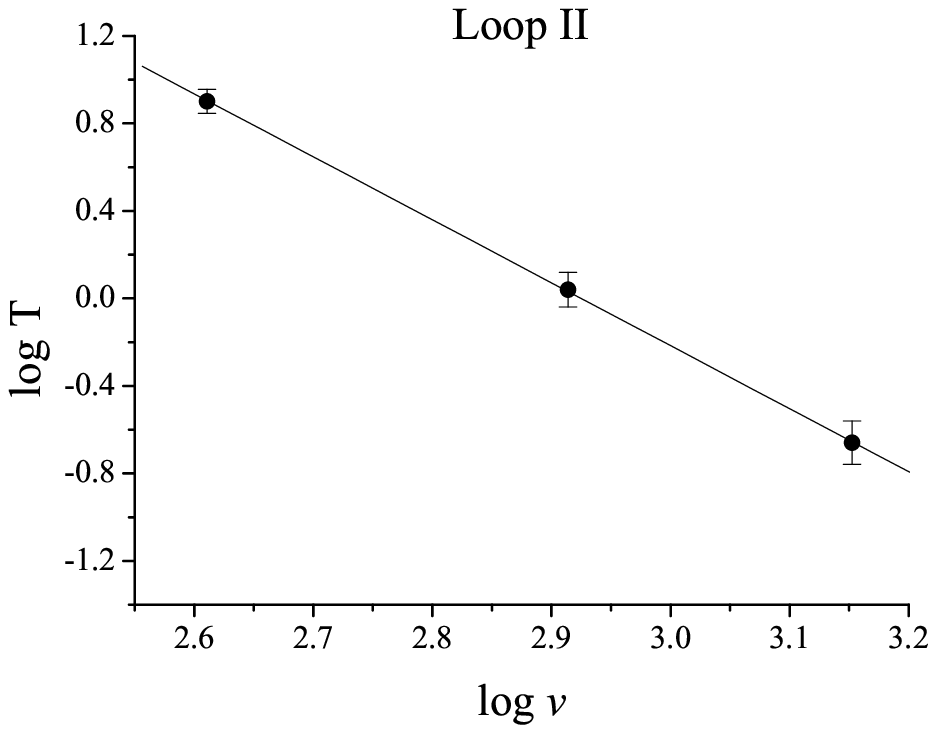}
\caption{The same as in Fig. $\ref{fig01}$, but for Loop II.}
\label{fig02}
\end{figure}

\begin{figure}
\centering
\includegraphics[width=0.45\textwidth]{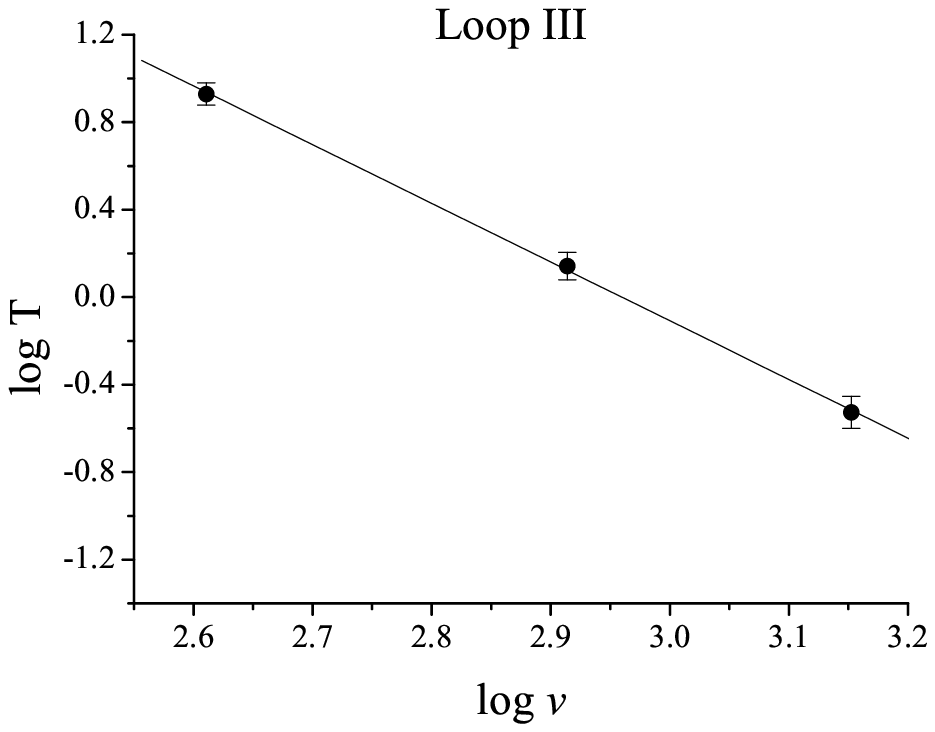}
\caption{The same as in Fig. $\ref{fig01}$, but for Loop III.}
\label{fig03}
\end{figure}

\begin{figure}
\centering
\includegraphics[width=0.45\textwidth]{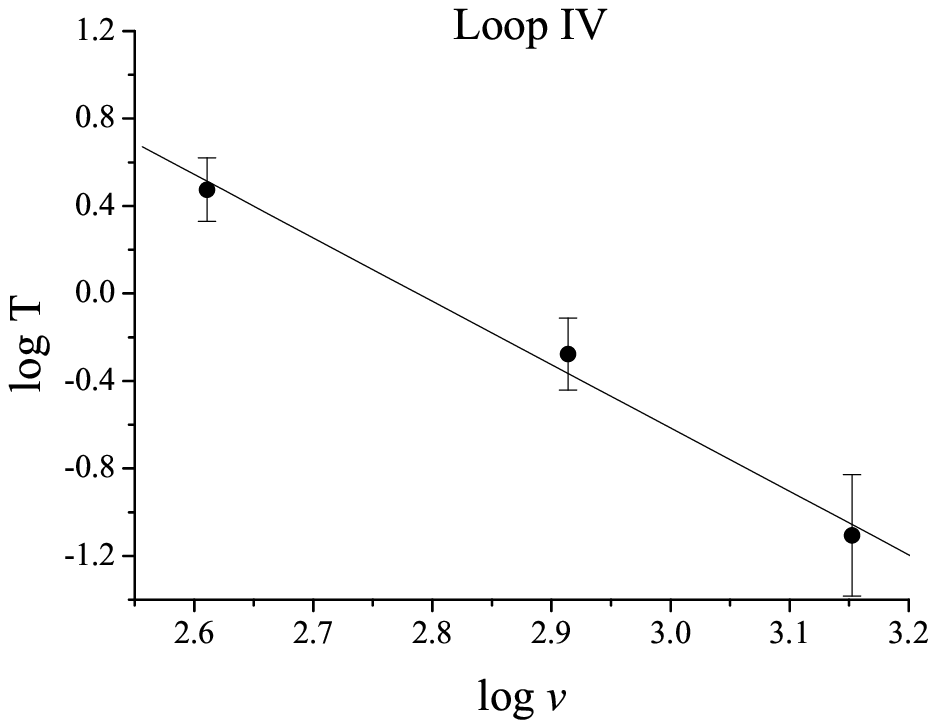}
\caption{The same as in Fig. $\ref{fig01}$, but for Loop IV.}
\label{fig04}
\end{figure}

\begin{figure}
\centering
\includegraphics[width=0.45\textwidth]{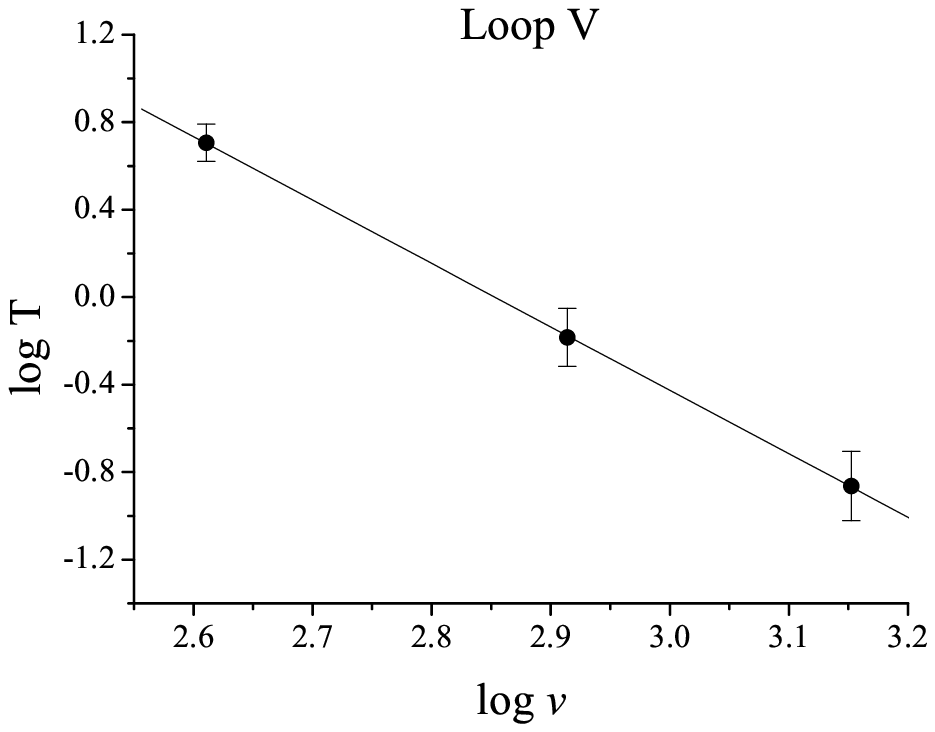}
\caption{The same as in Fig. $\ref{fig01}$, but for Loop V.}
\label{fig05}
\end{figure}

\begin{figure}
\centering
\includegraphics[width=0.45\textwidth]{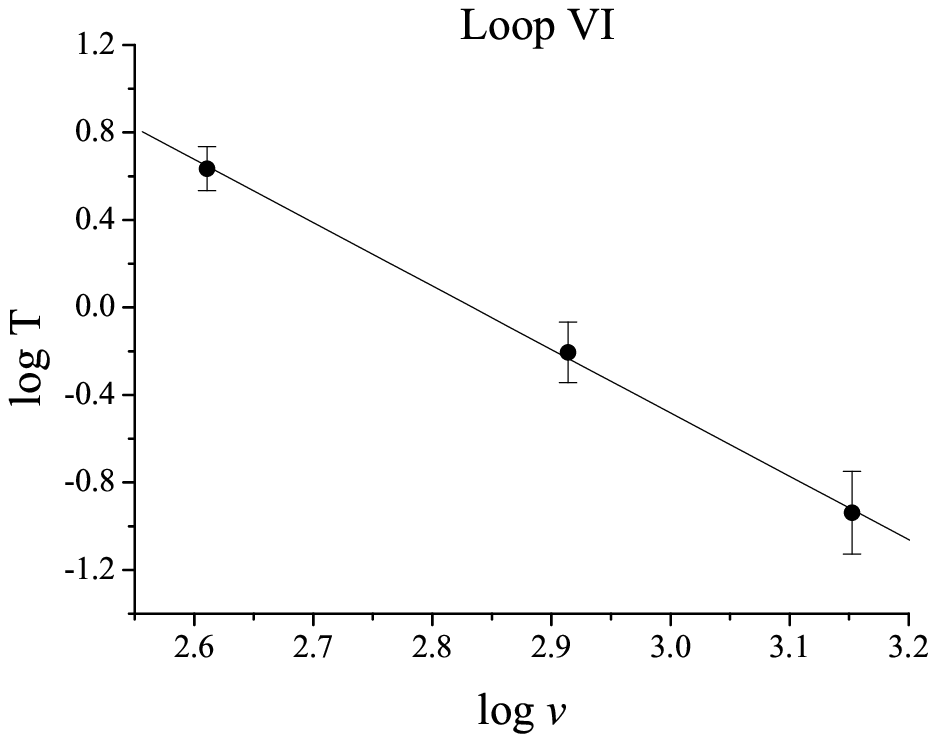}
\caption{The same as in Fig. $\ref{fig01}$, but for Loop VI.}
\label{fig06}
\end{figure}

Assuming that the radio loops are SNRs, they should follow the
$\Sigma - D$ (surface brightness -- diameter) relation for supernova
remnants. We have used the relation derived by \citet{case98} to
compute diameters of the Loops I-VI:

\begin{equation}
\Sigma _{1\mathrm{GHz}} = 2.82\cdot 10^{ {\rm -} 17} D^{ {\rm -}
2.41}
\end{equation}

\noindent and then for computing distances we use the relation:

\begin{equation}
r = D / (2 \sin \theta )
\end{equation}

\noindent with angular radii ($\theta )$ taken from \citet{jmt72}.
The results are given in Table \ref{tab05}. The largest SNRs that
\citet{case98} used to derive their $\Sigma - D$ relation had
diameters of about 100 pc, and 1-GHz brightnesses of about $3 \cdot
10^{-22}~W\,m^{-2}\,Hz^{-1}\,Sr^{-1}$. Hence, the application of
their relation to the Loops (with $\Sigma _{1\mathrm{GHz}} < 2.5
\cdot 10^{-22}~W\,m^{-2}\,Hz^{-1}\,Sr^{-1}$ from Table \ref{tab04})
contains an element of extrapolation.

\begin{table}
\centering
\caption{Spectral indices of radio loops, between 1420, 820 and 408 MHz.}
\begin{tabular}
{|l|c|}
\hline
label of the loop&
$\beta$ \\
\hline
\hline
\textbf{Loop I}&
 2.74 $\pm $ 0.08 \\
\hline
\textbf{Loop II}&
 2.88 $ \pm $ 0.03 \\
\hline
\textbf{Loop III}&
2.68 $ \pm $ 0.06 \\
\hline
\textbf{Loop IV}&
 2.90 $ \pm $ 0.28 \\
\hline
\textbf{Loop V}&
 3.03 $ \pm $ 0.15 \\
\hline
\textbf{Loop VI}&
 2.90 $ \pm $ 0.09 \\
\hline
\end{tabular}
\label{tab03}
\end{table}

\begin{table}
\centering
\caption{Brightnesses
(10$^{-22}$~W\,m$^{-2}$\,Hz$^{-1}$\,Sr$^{-1}$) reduced to 1000
MHz.}
\begin{tabular}
{|l|c|c|c|}
\hline
label &
\textbf{1420 MHz}&
\textbf{820 MHz}&
\textbf{408 MHz} \\
of the&
reduced to&
reduced to&
reduced to \\
loop&
\textbf{1000 MHz}&
\textbf{1000 MHz}&
\textbf{1000 MHz} \\
\hline
\hline
\textbf{Loop I}&
2.19 $\pm $ 0.72&
2.40 $\pm $ 0.47&
2.22 $\pm $ 0.42 \\
\hline
\textbf{Loop II}&
1.85 $\pm $ 0.45&
1.90 $\pm $ 0.34&
1.85 $\pm $ 0.26 \\
\hline
\textbf{Loop III}&
2.32 $\pm $ 0.62&
2.50 $\pm $ 0.38&
2.36 $\pm $ 0.40 \\
\hline
\textbf{Loop IV}&
0.66 $\pm $ 0.74&
0.91 $\pm $ 0.39&
0.68 $\pm $ 0.42 \\
\hline
\textbf{Loop V}&
1.15 $\pm $ 0.46&
0.97 $\pm $ 0.34&
1.14 $\pm $ 0.24 \\
\hline
\textbf{Loop VI}&
0.98 $\pm $ 0.41&
1.08 $\pm $ 0.34&
0.98 $\pm $ 0.26 \\
\hline
\end{tabular}
\label{tab04}
\end{table}

The smallest values of $\beta$ occur in the coldest parts of the sky
and the spectral index in the vicinity of Loops I and III is steeper
than the spectrum of the background emission. A qualitative
explanation of the steepening of the spectrum of emission apparently
associated with the Loops I and III has been put forward in terms of
diffusive shock-front acceleration \citep{laws87}: there is
compression of the interstellar cosmic ray electrons and magnetic
field behind a cooling SNR shock front. The resulting map of
\citet{reic88a} shows variations of the spectral index between
$\beta = 2.3$ and $\beta = 3.0$ and that the areas with steepest
spectra are those of the North Polar Spur and of Loop III.

When comparing the derived spectral indices with those of the
Galactic background (about 2.5 between 408 and 38 MHz, given in
\citet{laws87}), we can see from Table \ref{tab03} that these radio
loops have steeper, but not much steeper, spectra (about
$2.7 < \beta < 2.9$), which provides evidence for the loops being
very old SNRs.

The values for the brightness-temperature spectral indices of the
loops are rather steep (about $2.7 < \beta < 3.0 $). This is at the
high end of the spectral index distribution for SNRs in the range l
= [$360^{o}, 180^{o}$] and $D < 36\ pc $ given by \citet{clar76},
which has a mean T$_\mathrm{b}$ spectral index of 2.45. This might
be evidence for the steepening of SNR spectra with age, although
some authors (see e.g. \citet{clar76}) think there is no indication
that the spectral index is correlated with any other features of the
SNR.

\begin{table*}
\centering
\caption{Diameters (pc) and distances (pc) of the radio
loops derived from the $\Sigma - D$ relation given by
\citet{case98}. Diameters and distances are calculated from
brightnesses at 1420, 820 and 408 MHz respectively, when reduced to
1 GHz. In the last two columns there are average values of the
diameters and distances of the radio loops.}
\begin{tabular}
{|l|c|c|c|c|c|c|c|c|} \hline label of &
\multicolumn{2}{|c|}{\textbf{1420 MHz}} &
\multicolumn{2}{|c|}{\textbf{820 MHz}} &
\multicolumn{2}{|c|}{\textbf{408 MHz}} &
\multicolumn{2}{|c|}{\textbf{average}}  \\
\cline{2-9} the loop & diameter& distance& diameter& distance&
diameter& distance& diameter& distance \\
\hline \hline \textbf{Loop I}& 132 $\pm $ 20& 77 $\pm $ 14& 127
$\pm $ 19& 74 $\pm $ 13& 131 $\pm $ 20&
77 $\pm $ 14& 130 $\pm $ 20& 76 $\pm $ 14 \\
\hline \textbf{Loop II}& 141 $\pm $ 22& 97 $\pm $ 18& 140 $\pm $
21& 97 $\pm $ 18& 141 $\pm $ 22&
98 $\pm $ 18& 141 $\pm $ 22 & 97 $\pm $ 18 \\
\hline \textbf{Loop III}& 129 $\pm $ 20& 115 $\pm $ 24& 125 $\pm $
19& 112 $\pm $ 23& 128 $\pm $ 20&
115 $\pm $ 23& 127 $\pm $ 22 & 114 $\pm $ 24 \\
\hline \textbf{Loop IV}& 218 $\pm $ 33& 325 $\pm $ 67& 190 $\pm $
29& 283 $\pm $ 58& 214 $\pm $ 32&
319 $\pm $ 65& 207 $\pm $ 32 & 309 $\pm $ 64 \\
\hline \textbf{Loop V}& 172 $\pm $ 26& 91 $\pm $ 16& 185 $\pm $
28& 98 $\pm $ 17& 173 $\pm $ 26&
91 $\pm $ 16& 177 $\pm $ 27 & 93 $\pm $ 17 \\
\hline \textbf{Loop VI}& 184 $\pm $ 28& 97 $\pm $ 17& 177 $\pm $
27& 93 $\pm $ 16& 184 $\pm $ 28&
97 $\pm $ 17& 182 $\pm $ 28 & 96 $\pm $ 17 \\
\hline
\end{tabular}
\label{tab05}
\end{table*}

\section{Conclusions}

In this paper we have calculated the temperatures and surface
brightnesses of the Galactic radio loops at 1420, 820 and 408 MHz.
We are supposing all radio loops to be SNRs (\citet{berk70};
\citet{berk71}; \citet{shkl71}; \citet{salt83}). Our results are
consistent with the SNR hypothesis and suggest that the radio loops
may have a SNR origin. We have used data from the northern part of
the radio survey at 1420 MHz \citep{reic86} and at 820 MHz
\citep{berk72}, and the all-sky survey at 408 MHz \citep{hasl82}.
Spectra (temperature versus frequency) have been plotted and these
are used to determine spectral indices for the main Galactic loops.

The effective sensitivity of the brightness temperatures are: 1.0 K
for 408 MHz, 0.2 K for 820 MHz, and about 50 mK T$_\mathrm{b}$ for
1420 MHz. The most precise measurements (the least relative errors)
are in case of 1420 MHz, so positions of the brightness temperature
contours of the loops are the most realistic for this frequency.
Brightnesses of the radio loops at 408 and 820 and 1420 MHz are in
good agreement when reduced to 1 GHz.

We have demonstrated the reality of Loop V and Loop VI. We have also
calculated the temperatures, surface brightnesses and spectral
indices.

The spectral indices that we derived can be compared with values
derived by \citet{reic88b} for Loops I and III. They discussed the
distribution of spectral indices of the Galactic radio continuum
emission between 1420 and 408 MHz across the northern sky, as well
as the global properties of the Galactic spectral index variations.
Our results for spectral indices can also be compared with
\citet{berk73} for Loops I-IV. We note that our values for spectral
indices are in between the corresponding values for Loops I and III
given in papers \citet{reic88b} and \citet{berk73}.

In this paper, we present the first radio continuum spectra for
the main radio loops, plus Loops V and VI, made using average
brightness temperatures at three different frequencies. We find
that good linear fits can be made to each of these, supplying
accurate spectral indices.

With calculated radio spectral indices, we were able to estimate
diameters of these loops and distances to them. That the radio
spectra of the loops are fitted rather well by power-low spectra
is consistent with a SNR origin for these features.
\newline \\
Fig. A1 in the Appendix shows temperature scales in K, for 1420, 820
and 408 MHz. Figs A2 -- A7 in the Appendix show Loops I-VI at 1420,
820 and 408 MHz, with two contours representing the temperatures
T$_\mathrm{min}$ and T$_\mathrm{max}$ as given in Table \ref{tab01}.
Borders between the three frequencies are somewhat different for
each loop probably due to small, random and systematic errors in the
calibrated data. Also, we suppose there are uncertainties of about
0.1 K in the border of spurs due to measurement errors, and there is
a tiny difference in the absorption of radio emission in the
interstellar medium at different wavelengths \citep{pach70}.


\appendix

\begin{figure}
\section[]{The Radio Loops and Spurs at 1420, 820 and 408 MH\lowercase{z}}
\centering
\includegraphics[width=0.49\textwidth]{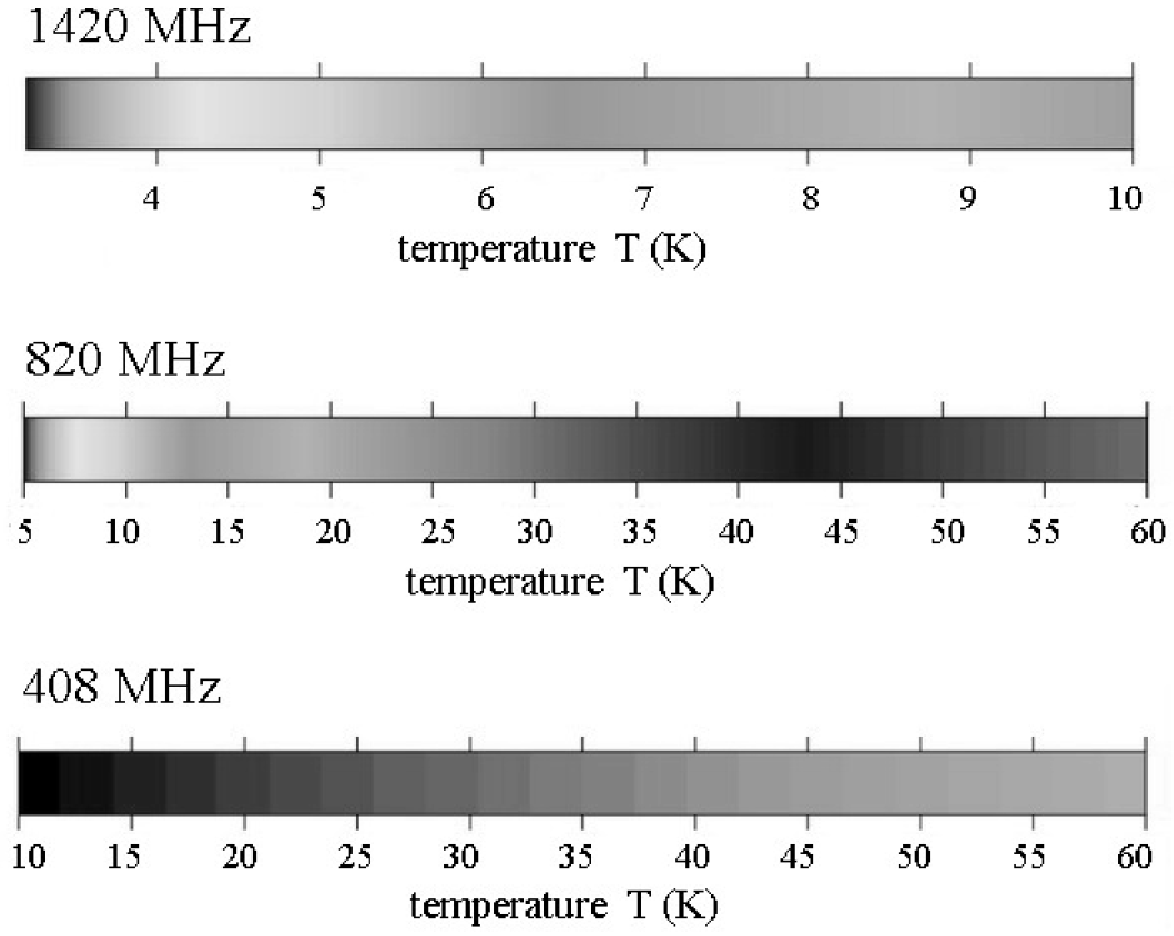}
\caption{Temperature scales for 1420, 820 and 408 MHz. All are given
in K and they are used for all pictures of the radio loops given
below.}
\label{figA1}
\end{figure}


\begin{figure}
\centering
\includegraphics[width=0.4\textwidth]{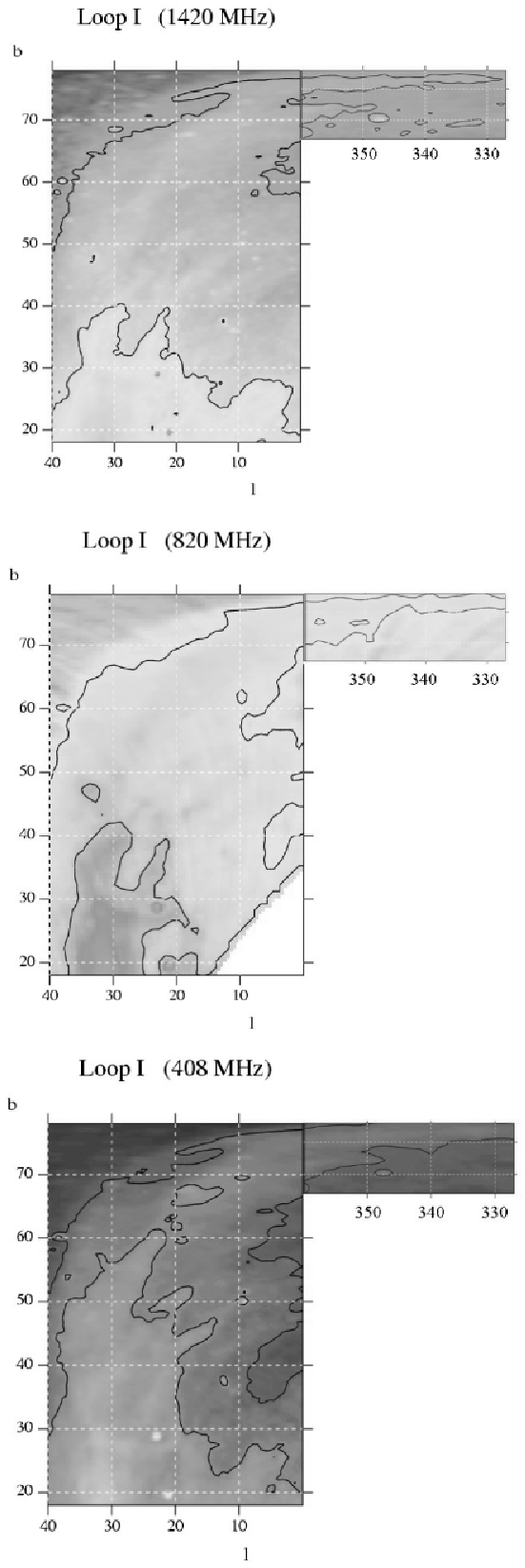}
\caption{The area of Loop I at 1420, 820 and 408 MHz, showing
contours of brightness temperature. This contains the part of the
North Polar Spur normal to the Galactic plane: l = [40$^\circ$, 0$^\circ$];
b = [18$^\circ$, 78$^\circ$] and its part parallel to the Galactic
plane: l = [360$^\circ$, 327$^\circ$]; b = [67$^\circ$, 78$^\circ$]. Two
contours are plotted, these representing the temperatures
T$_\mathrm{min}$ and T$_\mathrm{max}$, as given in Table
\ref{tab01}.}
\label{figA2}
\end{figure}

\clearpage

\begin{figure}
\centering
\includegraphics[width=0.49\textwidth]{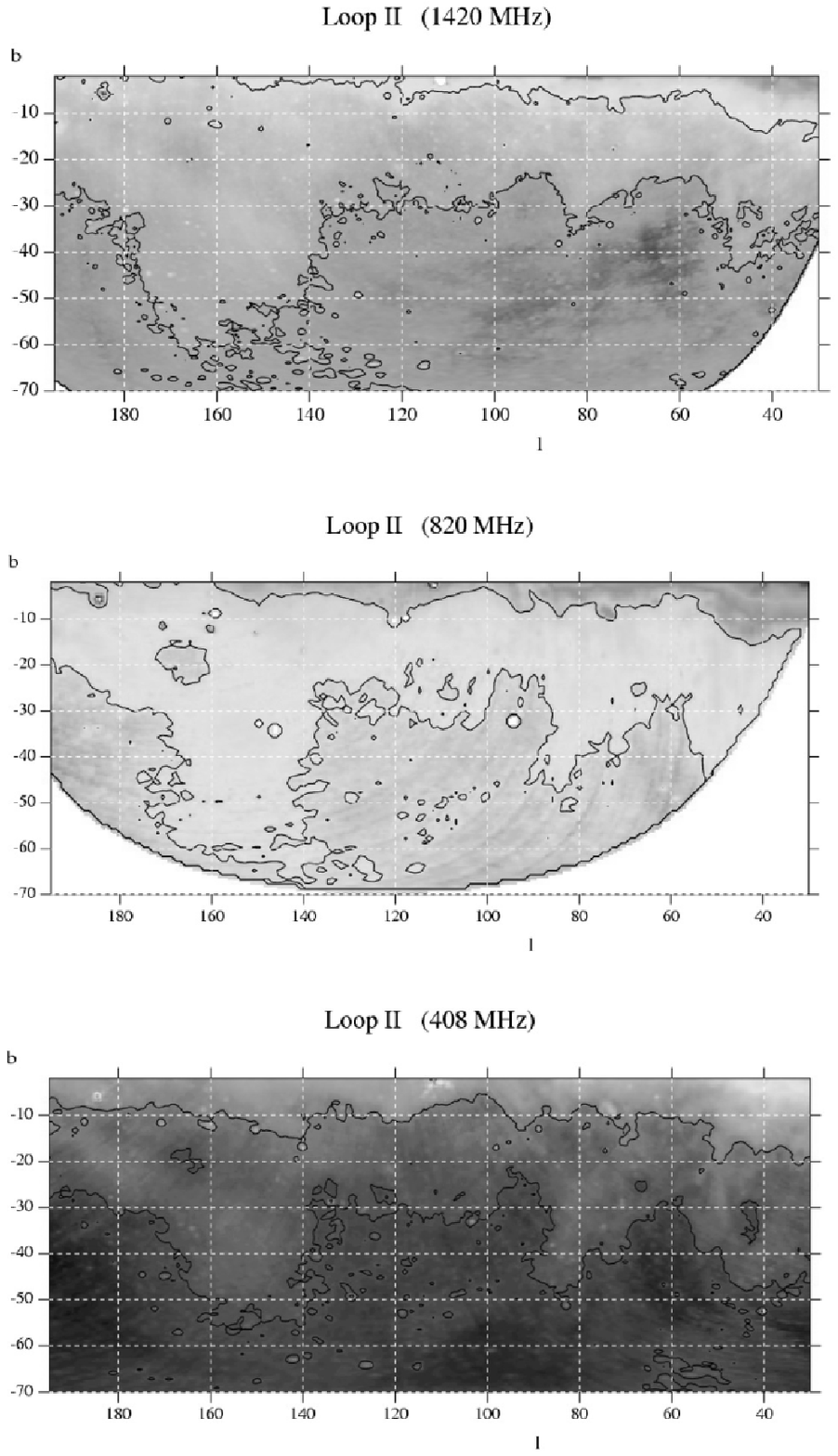}
\caption{The area of Loop II at 1420, 820 and 408 MHz, showing
contours of brightness temperature. The two contours plotted
represent the temperatures T$_\mathrm{min}$ and T$_\mathrm{max}$, as
given in Table \ref{tab01}. White areas in the pictures signify that
no data exist there at 1420 and 820 MHz. Spurs belonging to this
radio loop have positions: l = [57$^\circ$, 30$^\circ$]; b = [-50$^\circ$,
-10$^\circ$] for spur in Aquarius and l = [195$^\circ$, 130$^\circ$]; b =
[-70$^\circ$, -2$^\circ$] for spur in Aries.}
\label{figA3}
\end{figure}


\begin{figure}
\centering
\includegraphics[width=0.46\textwidth]{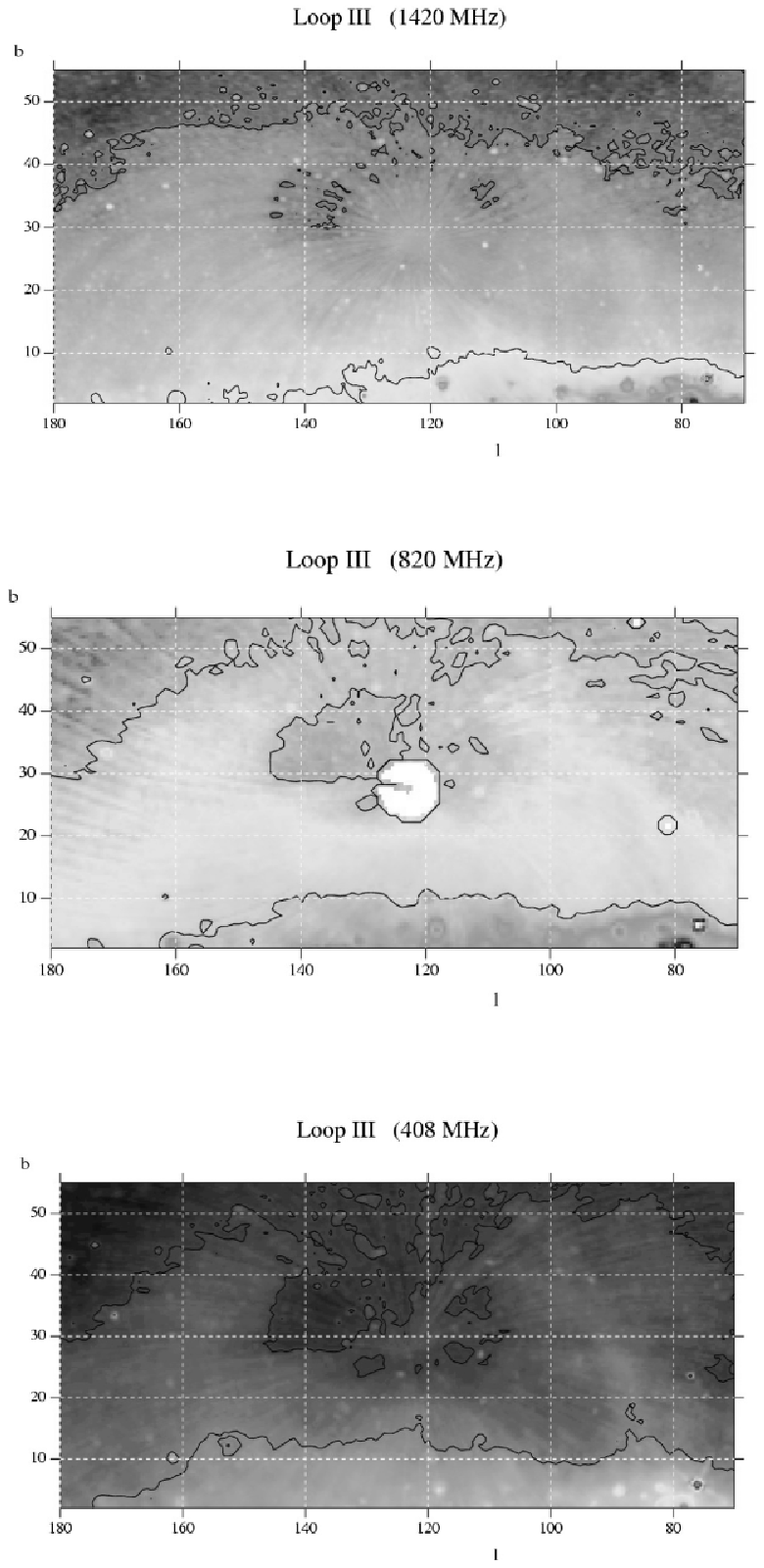}
\caption{The area of Loop III at 1420, 820 and 408 MHz, showing
contours of brightness temperature. The two contours plotted
represent the temperatures T$_\mathrm{min}$ and T$_\mathrm{max}$, as
given in Table \ref{tab01}. The white area in the picture for 820
MHz signifies that no data exist there at that frequency. Spurs
belonging to this radio loop have positions: l = [180$^\circ$,
135$^\circ$]; b = [2$^\circ$, 50$^\circ$] and l = [135$^\circ$, 110$^\circ$];
b = [40$^\circ$, 55$^\circ$] for first spur and l = [110$^\circ$,
70$^\circ$]; b = [6$^\circ$, 50$^\circ$] for second one.}
\label{figA4}
\end{figure}

\clearpage

\begin{figure}
\centering
\includegraphics[width=0.46\textwidth]{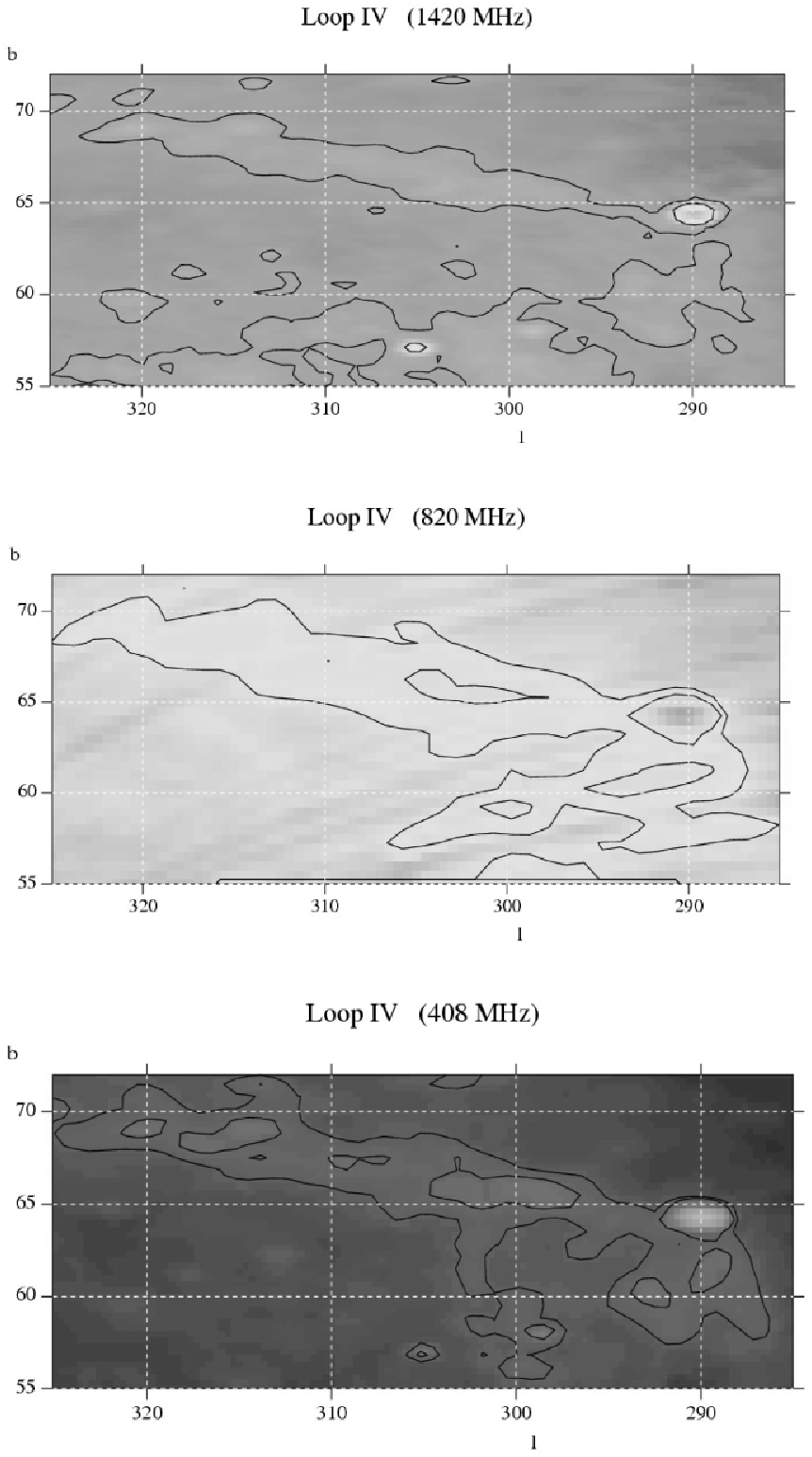}
\caption{The area of Loop IV at 1420, 820 and 408 MHz, showing
contours of brightness temperature. The two contours plotted
represent the temperatures T$_\mathrm{min}$ and T$_\mathrm{max}$, as
given in Table \ref{tab01}. This radio spur has position: l =
[330$^\circ$, 290$^\circ$]; b = [48$^\circ$, 70$^\circ$].}
\label{figA5}
\end{figure}


\begin{figure}
\centering
\includegraphics[width=0.49\textwidth]{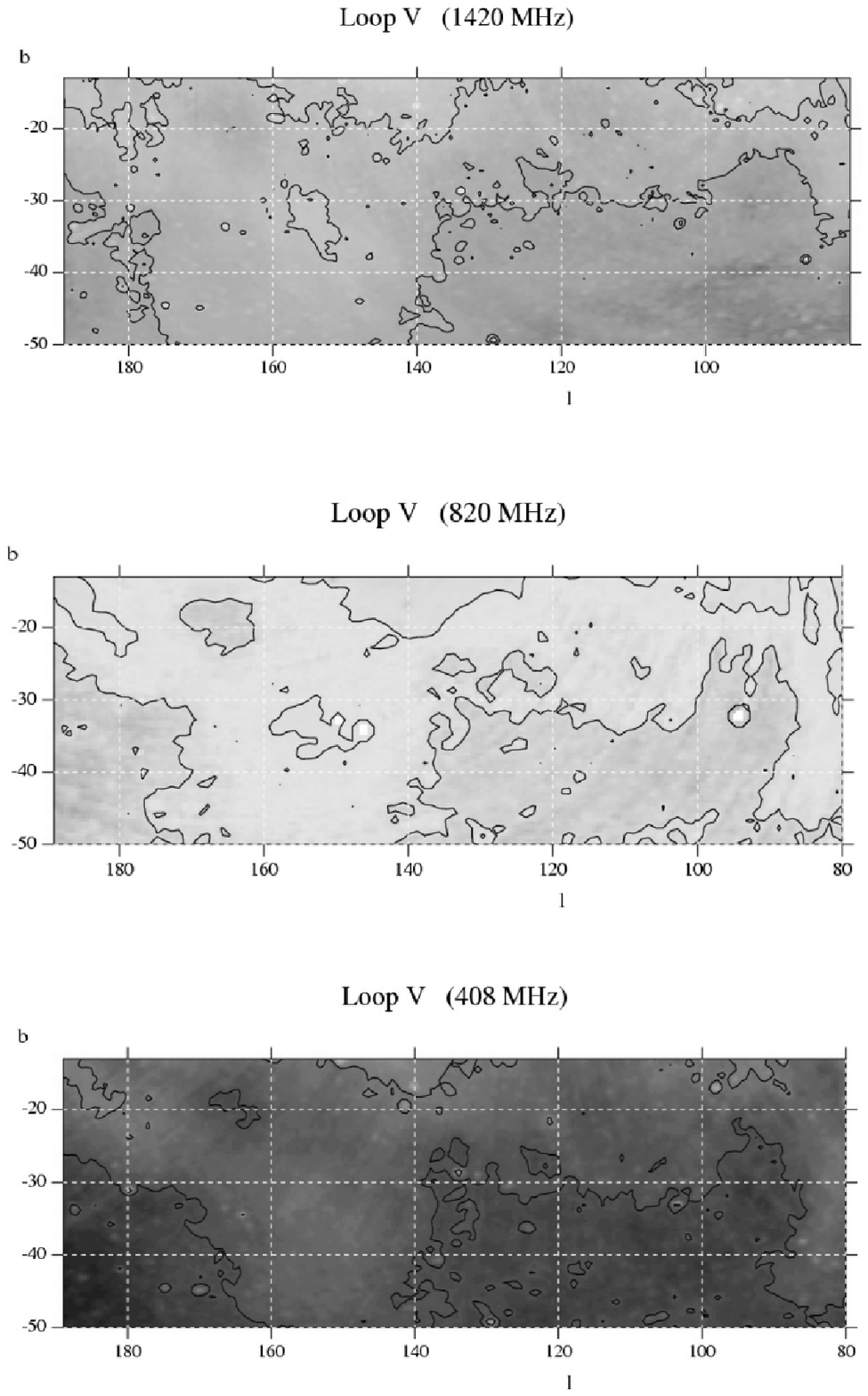}
\caption{The area of Loop V at 1420, 820 and 408 MHz, showing
contours of brightness temperature. The two contours plotted
represent the temperatures T$_\mathrm{min}$ and T$_\mathrm{max}$, as
given in Table \ref{tab01}. Spurs belonging to this radio loop have
positions: l = [189$^\circ$, 178$^\circ$]; b = [-25$^\circ$, -13$^\circ$]
for the spur in Taurus, l = [147$^\circ$, 133$^\circ$]; b = [-50$^\circ$,
-39$^\circ$] for the spur in Pisces and l = [90$^\circ$, 80$^\circ$]; b =
[-39$^\circ$, -24$^\circ$] for the spur in Pegasus.}
\label{figA6}
\end{figure}

\clearpage

\begin{figure*}
\centering
\includegraphics[width=0.8\textwidth]{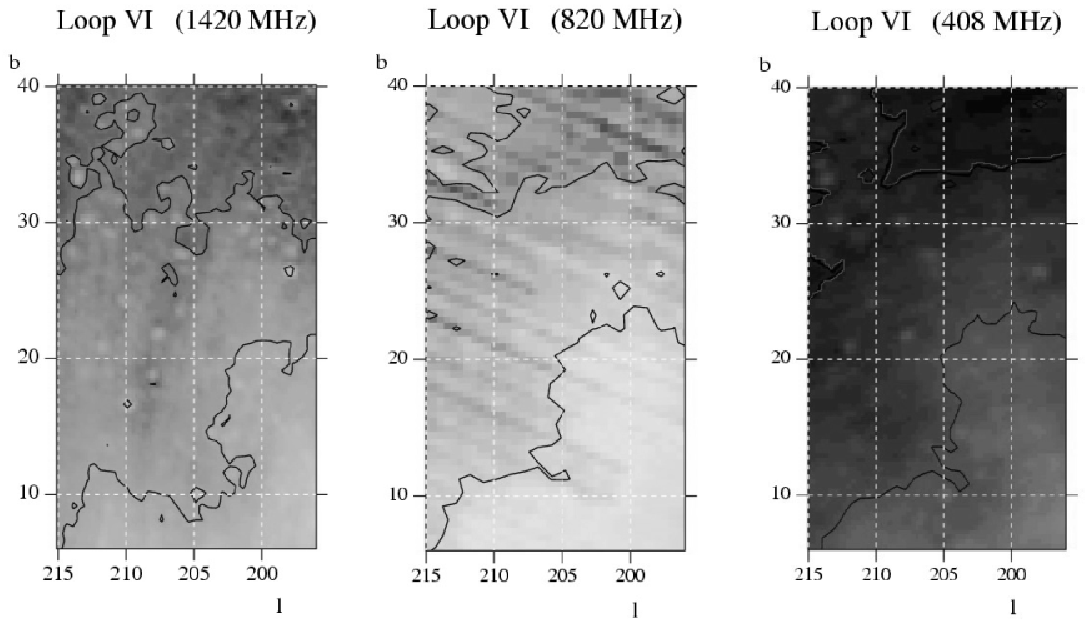}
\caption{The area of Loop VI at 1420, 820 and 408 MHz, showing
contours of brightness temperature. The two contours plotted
represent the temperatures T$_\mathrm{min}$ and T$_\mathrm{max}$, as
given in Table \ref{tab01}. Spurs belonging to this radio loop have
positions: l = [215$^\circ$, 205$^\circ$]; b = [29$^\circ$, 40$^\circ$] for
the spur in Leo and l = [207$^\circ$, 196$^\circ$]; b = [10$^\circ$,
32$^\circ$] for the spur in Cancer.}
\label{figA7}
\end{figure*}

\label{lastpage}


\begin{thebibliography}{99}
\bibitem[\protect\citeauthoryear{Berkhuijsen}{1971}]
{berk71} Berkhuijsen E.M., 1971, A\&A, 14, 359

\bibitem[\protect\citeauthoryear{Berkhuijsen}{1972}]
{berk72} Berkhuijsen E.M., 1972, A\&AS, 5, 263

\bibitem[\protect\citeauthoryear{Berkhuijsen}{1973}]
{berk73} Berkhuijsen E.M., 1973, A\&A, 24, 143

\bibitem[\protect\citeauthoryear{Berkhuijsen et al.}{1970}]
{berk70} Berkhuijsen E.M., Haslam, C. G. T.,
Salter, C. J., 1970, Nat, 225, 364

\bibitem[\protect\citeauthoryear{Berkhuijsen et al.}{1971}]
{berk71a} Berkhuijsen E.M., Haslam, C. G. T., Salter, C. J., 1971,
A\&A, 14, 252

\bibitem[\protect\citeauthoryear{Case \& Bhattacharya}{1998}]
{case98} Case, G. L., Bhattacharya, D., 1998, ApJ, 504, 761

\bibitem[\protect\citeauthoryear{Clark \& Caswell}{1976}]
{clar76} Clark, D. H., Caswell, J. L., 1976, MNRAS, 174, 267

\bibitem[\protect\citeauthoryear{Haslam et al.}{1964}]
{hasl64} Haslam, C. G. T., Large, M. I., Quigley, M. J. S., 1964,
MNRAS, 127, 273

\bibitem[\protect\citeauthoryear{Haslam et al.}{1982}]
{hasl82} Haslam, C. G. T., Salter, C. J., Stoffel, H., Wilson, W.
E., 1982, A\&AS, 47, 1

\bibitem[\protect\citeauthoryear{Large et al.}{1962}]
{larg62} Large, M. I., Quigley, M. J. S., Haslam, C. G. T., 1962,
MNRAS, 124, 405

\bibitem[\protect\citeauthoryear{Large et al.}{1966}]
{larg66} Large, M. I., Quigley, M. J. S., Haslam, C. G. T., 1966,
MNRAS, 131, 335

\bibitem[\protect\citeauthoryear{Lawson et al.}{1987}]
{laws87} Lawson, K. D., Mayer, C. J., Osborne, J. L., Parkinson,
M. L., 1987, MNRAS, 225, 307

\bibitem[\protect\citeauthoryear{McKee \& Ostriker}{1977}]
{mcke77} McKee, C. F., Ostriker, J. P., 1977, ApJ, 218, 148

\bibitem[\protect\citeauthoryear{Milogradov-Turin}{1972}]
{jmt72}Milogradov-Turin, J., 1972, M. Sc. Thesis, Univ. of
Manchester

\bibitem[\protect\citeauthoryear{Milogradov-Turin}{1982}]
{jmt82}Milogradov-Turin, J., 1982, Ph. D. Thesis, Univ. of
Belgrade

\bibitem[\protect\citeauthoryear{Milogradov-Turin \&
Nikoli\'c}{1995a}] {jmt95a} Milogradov-Turin, J., Nikoli\'c, S.,
1995a, BA Belgrade, 151, 7

\bibitem[\protect\citeauthoryear{Milogradov-Turin \&
Nikoli\'c}{1995b}] {jmt95b} Milogradov-Turin, J., Nikoli\'c, S.,
1995b, BA Belgrade, 152, 11

\bibitem[\protect\citeauthoryear{Milogradov-Turin \&
Smith}{1973}] {jmt73} Milogradov-Turin, J., Smith F. G., 1973,
MNRAS, 161, 269

\bibitem[\protect\citeauthoryear{Milogradov-Turin \& Uro\v{s}evi\'c}{1997}]
{jmt97} Milogradov-Turin, J., Uro\v{s}evi\'c, D., 1997, BA
Belgrade, 155, 41

\bibitem[\protect\citeauthoryear{Pacholczyk}{1970}]
{pach70} Pacholczyk, A. G. 1970, Radio Astrophysics, Freeman, W.
H. and Company, San Francisco

\bibitem[\protect\citeauthoryear{Quigley \& Haslam}{1965}]
{quig65} Quigley, M. J. S., Haslam, C. G. T., 1965, Nat, 208,
 741

\bibitem[\protect\citeauthoryear{Reich \& Reich}{1986}]
{reic86} Reich, P., Reich, W., 1986, A\&AS, 63, 205

\bibitem[\protect\citeauthoryear{Reich \& Reich}{1988a}]
{reic88a} Reich, P., Reich, W., 1988a, A\&AS, 74, 7

\bibitem[\protect\citeauthoryear{Reich \& Reich}{1988b}]
{reic88b} Reich, P., Reich, W., 1988b, A\&A, 196, 211

\bibitem[\protect\citeauthoryear{Reich \& Steffen}{1981}]
{reic81} Reich, W., Steffen, P., 1981, A\&A , 93, 27

\bibitem[\protect\citeauthoryear{Salter}{1970}]
{salt70} Salter, C. J., 1970, Ph. D. Thesis, Univ. of Manchester

\bibitem[\protect\citeauthoryear{Salter}{1983}]
{salt83} Salter, C. J., 1983, BAS of India , 11, 1

\bibitem[\protect\citeauthoryear{Shklovskii \& Sheffer}{1971}]
{shkl71} Shklovskii, I. S, Sheffer, E. K., 1971, Nat, 231, 173

\bibitem[\protect\citeauthoryear{Webster}{1974}]
{webs74} Webster, A. S., 1974, MNRAS, 166, 355


\end{thebibliography}
\end{document}